\documentclass[aps,pre,twocolumn]{revtex4-1}

\usepackage{dcolumn}
\usepackage{amsmath}
\usepackage{amssymb}
\usepackage{bm}
\usepackage{graphicx}
\usepackage{epsf,epsfig}
\usepackage{hyperref}
\usepackage{color}

\begin{document}
\title{How does an embedded spheroid affect the rigidity of extracellular matrix?}
\author{Amanda Parker$^{1,2}$ and J. M. Schwarz$^{1,3}$}
\affiliation{$^1$Department of Physics and BioInspired Syracuse, Syracuse University, Syracuse, NY 13244, USA\\
$^2$ SymBioSysm, Inc., Chicago, Illinois, 60601, USA\\
$^3$Indian Creek Farm, Ithaca, NY, 14850, USA}
\date{\today}

\begin{abstract}

\textit{In vitro} collagen networks and \textit{in silico} fiber network models are typically used to represent extracellular matrix in tissues. Such networks exhibit the phenomenon of strain-stiffening, or an increase in elastic modulus with increasing strain, both under isotropic expansion and shear. However, the deformations induced in an extracellular matrix environment in the presence of a cellular aggregate are more complex, due to the irregularity of the tissue-environment interface, the mechanisms of force transmission between the tissue and the environment, and the rheology of the tissue itself. Therefore, using a two-dimensional vertex model of a tissue coupled to a surrounding spring network model, both of which can undergo rigidity transitions, we investigate the effects of a cellular aggregate on the rigidity of its environment. We find that the network's rigidity transition alone is sensitive to tissue size, mechanical properties, and surface tension. This sensitivity can, in part, be analytically estimated using a mean-field constraint counting approach to arrive at an effective spring network coordination number to determine how the network rigidity transition location shifts in the presence of the tissue spheroid. Moreover, we find that it is energetically favorable to create a ring of high-tension boundary cells in the tissue spheroid as the spring network rigidifies, as opposed to creating a string of high-tension cells through the bulk. We also find that increasing interfacial tension of the tissue spheroid facilitates rigidity in the spring network. In sum, our numerical and analytical results help reveal the complex mechanical interplay between a tissue spheroid and its surrounding environment. 


\end{abstract}

\maketitle

\section{Introduction}

Biological tissue is composed of cells and extracellular matrix (ECM). The ECM is a filamentous mesh that surrounds the cells. It provides structural support for cells, helps regulate cell-cell adhesion, and facilitates cell migration \cite{Frantz_2010, Tang_2020, Bonnans_2014}. To do so, collagen, the primary component of ECM in vertebrates, attaches to cells via collagen-binding integrins, which facilitate bonds between the cell cytoskeleton and ECM collagen molecules. Although there are many types of cell-matrix adhesions, differing in the types of cellular and ECM components to which they connect, adhesions to collagen, particularly to collagen type I, are the strongest and most efficient at transmitting forces over long ranges~\cite{Frantz_2010,THEOCHARIS_2016,Tang_2020}. Collagen-binding integrins allow cells to exert forces on the ECM, whose mechanical response acts as a signal to change cell properties, tissue behavior, and, therefore, function~\cite{Yusko_2014,  Balaji_2019, Lo_2000, Lui_2017, Humphrey_2014, Frantz_2010}. 

Given the importance of collagen in the ECM, {\it in vitro} and {\it in silico} collagen networks are well-studied, particularly in terms of their microstructure and mechanical properties~\cite{Stein2011micromechanics,Lee2014,Kim2017stress,Dong_2017,Jansen_2018,Sun2021mechanics}. A collagen network exhibits the phenomenon of strain-stiffening at large strain, in which its shear modulus increases with increasing shear strain, indicating a nonlinear mechanical material~\cite{Storm2005nonlinear,Motte2013strain}. Various biophysical mechanisms driving the strain-stiffening have been proposed~\cite{Sharma_2016,Rens2018micromechanical,Merkel_2019}. Researchers have also explored the mechanical effects of individual cells embedded in collagen networks~\cite{Wolf_2009, Schor_1980, Hall_2016, Liang2016heterogeneous,Han2018cell,Kim_2018,Zheng_2019}. In addition to individual cells remodeling collagen fibers, collagen networks, and related fibrous networks, transform from compression softening materials to compression stiffening materials in the presence of cells~\cite{Van2019emergence,Gandikota2020loops,engstrom2019compression}. The combination of cells and collagen indeed results in complex mechanical behavior which is helping grow our understanding of healthy tissue function.  

In addition, a quantitative understanding of how ECM influences cells in a tissue is helping to elucidate observed abnormalities in the mechanical properties of cancerous tumors \cite{Walker_2018, GIUSSANI_2015, Acerbi_2015, Leight_2017, McKenzie_2018}. In particular, understanding how the mechanical state of the ECM, represented by a collagen network in many cases, influences the migratory behavior of tumor cells breaking out of a tumor spheroid has gained intense academic interest from both biologists and physicists \cite{Kopanska_2016, Suh_2019, Huang2020, Baskaran_2020,Pandey2023,van2024emt}. In fact, we recently investigated in two dimensions, numerically and analytically, how coupling a simple spring network to a spheroid model alters the spheroid's phase diagram. We focused primarily on limits of our model in which the environment was either rigid or floppy and examined the resulting spheroid states. We found that changes in the mechanical and morphological states of the spheroid could indeed be induced by altering the mechanical state of the environment \cite{Parker_2020}. Here, we now investigate the converse question: how does the mechanical state of the spheroid affect the mechanics of the extracellular matrix?  In exploring this, we study how the rigidity transition in the spring network is influenced by tissue spheroid size and tissue spheroid properties, such as fluidity and the shapes of the tissue spheroid's constituent cells.

In Section \ref{sec_model}, we describe our model and define quantities and methods used throughout. In Section \ref{sec_results}, we vary the relative size of the embedded tissue and the balance of cell-cell adhesion and contractility of its cells, and measure the transition point of the surrounding network. We then look more broadly at changes to the nature of the network's rigidity transition, due to the presence of the tissue. We then add a tissue-network interfacial tension term to our Hamiltonian, and observe the effect its magnitude has on the rigidity of the network.

\section{Model and methods} \label{sec_model}

\subsection{Hamiltonian}

\begin{figure}
    \centering
    \includegraphics[width=0.43\textwidth]{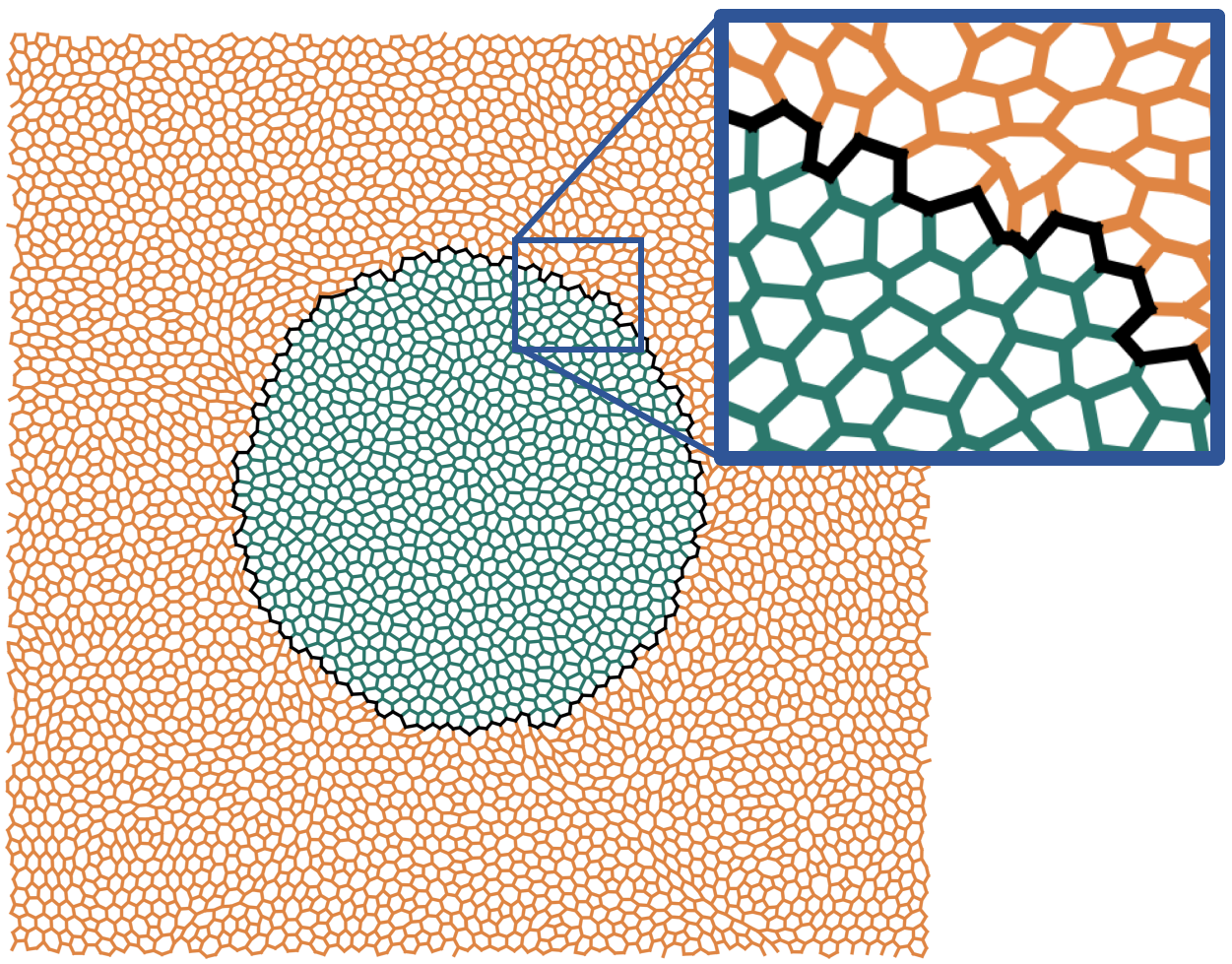}
    \caption{\textit{Model illustration.} A snapshot of a system configuration in which $N_{box}=2000$ total polygons. The tissue (region of green-edged polygons) takes up roughly 20\% of the total box area and therefore contains approximately $N_{tissue}=400$ cells. The spring network (orange edges) spans the rest of the system. The edges on the interface between the tissue and the spring network are shown in black.}
    \label{fig_model}
\end{figure}{}

To model a cross-section of a tissue spheroid embedded in extracellular matrix (ECM), we start with a two-dimensional Voronoi tiling of a periodic box. We then designate a circular region of polygons to represent the tissue, while all polygons whose centers fall outside the circle make up the ECM. A typical configuration is shown in Fig. \ref{fig_model}. 

The tissue region is described using a standard vertex model, in which each polygon represents a cell and has energy
\begin{equation}
    E_{cell,\alpha} = K_{A,\alpha}(A_\alpha-A_{0,\alpha})^2 + K_{P,\alpha}(P_\alpha-P_{0,\alpha})^2,
\end{equation}
where $A_\alpha$ and $P_\alpha$ are the area and perimeter of cell $\alpha$, which are functions of the vertex positions \cite{Farhadifar_2007, Bi_2015}.

The surrounding ECM region is described by a simple spring-network model, where the energy of each edge is
\begin{equation}
    E_{spring,ij} = \frac{1}{2}K_{ij}(L_{ij}-L_{0,ij})^2,
\end{equation}
and $L_{ij}$ is the length of edge $ij$. Note that this energy contribution does not contain bending energies as we have decided to explore the more dramatic rigidity transition in the spring network as it is isotropically strained. Potential implications of additional three-body interactions are discussed in the final section. 

Assuming all coupling constants are the same for each cell and spring, the dimensionless, total energy of the system is given by 
\begin{multline}
    e_{total} = \sum_{cells, \alpha} \Big[(a_{\alpha} - 1)^{2} + k_{p} (p_{\alpha} - p_{0})^{2} \Big]\\ 
     + \frac{1}{2} k_{sp} \sum_{springs, ij }  (l_{ij} - l_{0})^{2}
 \label{eq_nondim_energy_total}
 \end{multline}
where the first sum is over all cells $\alpha$ and the second is a sum over all springs $ij$.

The parameter $p_{0}$, which we call the ``preferred shape index'' of a cell, captures the balance of cell-cell adhesion and cell contractility, and controls the rigidity in the vertex model for bulk tissue. Similarly, in a homogeneous, under-constrained spring network system, a control parameter for rigidity is the ``preferred'' length of the springs, $l_0$. In both of these systems, there is a critical value of the control parameter, above which the system is floppy, or fluid-like, and below which the system is rigid, or solid-like \cite{Merkel_2019, Bi_2015}. For bulk tissue, the critical preferred shape value is $p^*_0 \approx 3.81$, and for bulk spring network, the critical preferred spring length value is $l^*_0 \approx 0.63$, though these values depend somewhat on simulation protocol \cite{Bi_2016, Sussman_2018, Wang_2019, Merkel_2019}.

\subsection{Equation of motion}

To simulate our system with the Hamiltonian in Eq. \ref{eq_nondim_energy_total}, we use a modified version of the open-source software package, cellGPU \cite{Sussman_2017}. cellGPU is a molecular dynamics program designed specifically for vertex- and cell-based models of confluent tissues, which we adapt to include spring-type edges and rules for tissue-spring interactions. 

For a fixed set of parameters, we update the positions of all vertices in the system according to the FIRE minimization algorithm \cite{Bitzek_2006}. Once the system has found an energy minimum, we measure and compute quantities of interest. One such quantity is the critical $l_0$ value, $l^*_0$, of the spring network region.

\subsection{Spring tension and the spring network transition point}

To extract the critical value, $l^*_{0}$, of our spring network tuning parameter, $l_{0}$, we first define the participation ratio, $\psi$, of bond tensions, as done previously \cite{Arzash_2020}. Briefly, we define the tension in spring $ij$ as $T_{ij} = k_{sp}(l_{ij} - l_0)$ and choose a tension threshold below which we consider the tension to be effectively zero: $T_{thresh} = 1 \times 10^{-8}$. We compute $\psi$ for a configuration as the ratio of the number of springs with tension above the threshold to the number of total springs.

For a given initial configuration, we vary $l_0$ of the spring network and measure the participation ratio, $\psi$, at each value. We find that $\psi$ transitions from being $\approx 1$ to $\approx 0$ across $l_0$ values. To estimate the critical value, $l^*_{0}$, at which the system transitions, we determine the point at which the absolute value of the slope of $\psi$ vs. $l_{0}$ is the greatest. We then repeat this for a series of initial configurations and compute the mean and standard error of the extracted $l^*_{0}$ values.

\subsection{Tissue cell edge tension and tissue rheology}

As mentioned earlier, the vertex model used in this study to represent the tissue region undergoes a rigidity transition as a function of the preferred cell shape index, $p_0$. There are many ways to quantify the fluidity of the tissue, including measuring the mean squared displacement or rearrangement rate of tissue cells at finite temperature, or, at zero temperature, measuring the energy barriers to undergoing cell rearrangements or the tissue shear modulus. All of these methods quantify the frustration of the tissue cells-- in the fluid-like phase, the cells are free to move and rearrange, because the energy barriers to doing so are low, while the opposite is the case in the rigid-like phase. Ultimately, this frustration is geometric, coming from whether or not cells can or cannot achieve their preferred shape indices. Therefore, another way to probe tissue rigidity is to examine the tensions along cell-cell contacts, defined as $T_{\alpha \beta} = k_p (p_{\alpha} - p_0) + k_p (p_{\beta} - p_0)$ for the edge shared between cells $\alpha$ and $\beta$. We utilize this measure to examine the spatial distribution of tensions throughout our entire composite system, at the onset of network rigidity.

\section{Results} \label{sec_results}

\subsection{The spring network transition point and effective coordination number} \label{sec_system_size_l0star}

To understand how the presence of the embedded tissue affects the spring network's rigidity diagram, we must first examine how simply changing the system size, $N_{box}$, and the fraction of the system taken up by tissue, $f_{tiss/box}$, affects the transition point of the spring network, $l^*_{0}$ (defined in Section \ref{sec_model}). To do this systematically, we first fix $p_{0}=3.71$ for the embedded tissue. This value is below the critical value, which means that in a homogeneous system, we would expect the tissue to act as a solid.

\begin{figure}
    \begin{center}
    \includegraphics[width=0.35\textwidth]{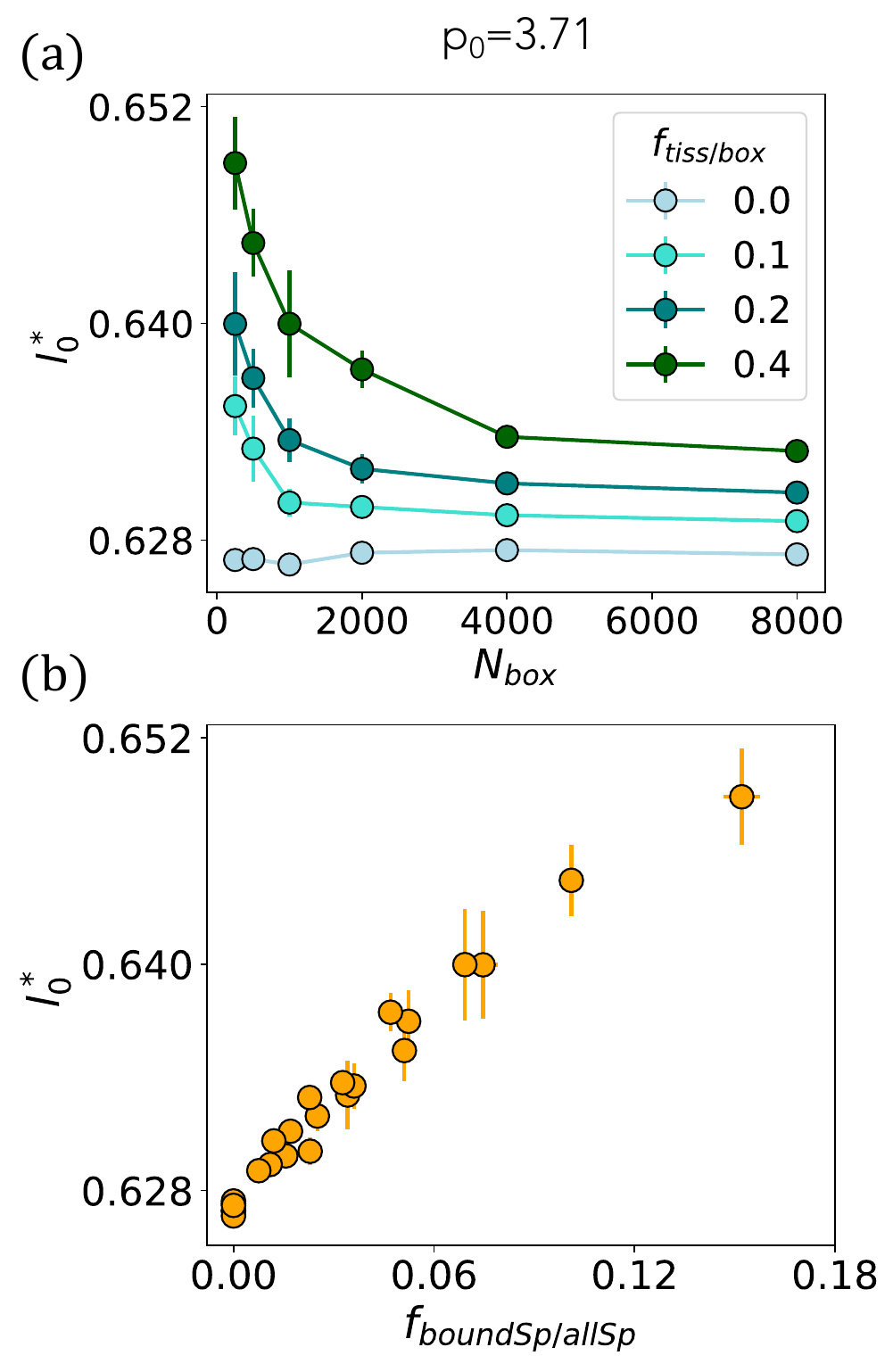}
    \caption{\textit{The dependence of $l^*_{0}$ on system size and tissue fraction, for a tissue with $p_0=3.71$.} (a) $l^*_{0}$ is plotted against the total system size for four values of the percentage of the box taken up by tissue. (b) For each data point in (a), we calculate the number of spring degrees of freedom connected to the boundary and plot this value (normalized by the total number of spring degrees of freedom) vs. $l^*_{0}$. 
    }
    \label{fig_p0_3p71_l0star_boxsize_fraction_DOF}
    \end{center}
\end{figure}{}

As previously reported, we find that the transition point of a spring network alone (with no embedded tissue) in a periodic box, does not depend strongly on system size \cite{Merkel_2019}. However, with embedded tissue, we find that $l^*_{0}$ does exhibit system-size-dependence. These results are shown in the upper plot of Fig. \ref{fig_p0_3p71_l0star_boxsize_fraction_DOF}, where we plot $l^*_{0}$ vs. $N_{box}$, for four values of $f_{tiss/box}$. For a fixed system size, increasing $f_{tiss/box}$ increases $l^*_{0}$, and does so more dramatically for smaller system sizes.


One feature of this data is that different combinations of system size and tissue fraction result in similar $l^*_{0}$ values. For example, $l^*_{0}(N_{box}=4000, f_{tiss/box}=0.4) \approx l^*_{0}(N_{box}=1000, f_{tiss/box}=0.2) \approx l^*_{0}(N_{box}=500, f_{tiss/box}=0.1)$. We also know, from Maxwell constraint-counting arguments and previous studies of strain-stiffening in spring networks, that the critical strain depends on the average connectivity of the network, which itself depends on the numbers of degrees of freedom and constraints \cite{Maxwell_1864, Sharma_2016}. We therefore hypothesize that because $l_0$ acts to strain the spring network, by changing the system size and tissue fraction we are effectively changing the connectivity of the spring network, and subsequently $l^*_{0}$, in a systematic way. 

To test this, we first compute the ratio of number of spring vertices attached to the tissue boundary to total number of spring vertices, which we call $f_{boundSp/allSp}$, for all of the data used to make the upper plot of Fig. \ref{fig_p0_3p71_l0star_boxsize_fraction_DOF}, and plot it against $l^*_{0}$. The results are shown in the lower plot of Fig. \ref{fig_p0_3p71_l0star_boxsize_fraction_DOF}. Upon doing this, we find a good collapse of our data, which supports our hypothesis that, for a fixed $p_{0}$ of the embedded tissue, $l^*_{0}$ is ultimately dependent on the degrees of freedom in the spring network.

We hypothesize that the collapse in Fig. \ref{fig_p0_3p71_l0star_boxsize_fraction_DOF} (b) is a consequence of changing the effective connectivity of the spring network, since, by replacing a portion of the network with tissue, we change the number of springs and impose a new boundary condition. To understand how the coordination number of the vertices in the spring network is related to the embedded tissue properties, we use a constraint-counting argument. 

For any system of $N_{vertices}$ and $N_{edges}$, where each edge is shared by exactly two vertices, the average coordination number (or average number of connections (edges) per vertex), $\langle z \rangle$, can be defined as 
\begin{equation}
    \langle z \rangle = \frac{2 N_{edges}}{N_{vertices}}.
 \label{eq_avg_coord_num_01}
\end{equation}
In our system, every vertex is connected by exactly three edges to exactly three neighbor vertices, and we use periodic boundary conditions, which means that the coordination number for the entire system is always exactly three. However, the spring network region is bounded on one side by the embedded tissue, and therefore computing $\langle z \rangle$ for the spring network alone requires considering the effect on spring-type degrees of freedom and constraints at the tissue-ECM interface.

\begin{figure}
    \centering
    \includegraphics[width=0.3\textwidth]{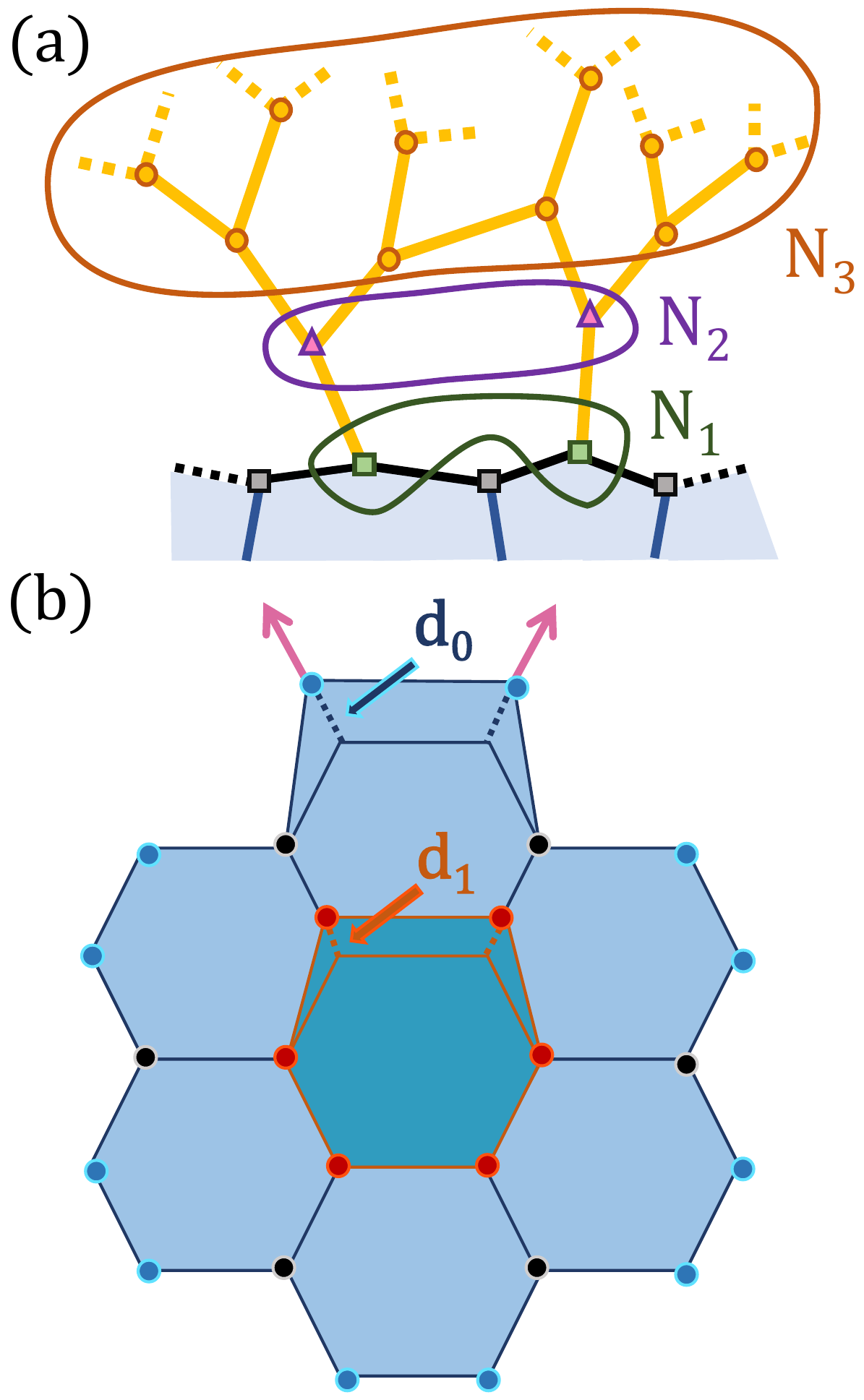}
    \caption{\textit{Illustrations of toy models.} (a) A representation of a portion of the tissue-ECM boundary. We divide the vertices that potentially contribute to the spring network's degrees of freedom into three sets: $N_1$ boundary-type vertices that are connected to springs (green squares), $N_2$ spring-type vertices that are connected to the boundary (pink triangles), and $N_3$ other spring-type vertices (yellow circles). (b) A representation of a tissue embedded in ECM. The tissue is constructed from seven total hexagons-- six on the boundary and one in the bulk. The outer vertices which connect to springs (blue circles) are displaced by $d_0$, which theoretically depends on the spring network tension, and the bulk vertices (red circles) compensate by displacing an amount $d_1$.}
    \label{fig_models_coordination_number_penetration_depth}
\end{figure}

We will consider the total numbers of three types of vertices in our constraint-counting of the spring network: the number of vertices on the boundary of the tissue which are connected to springs, $N_{1}$, the number of vertices connected by a spring to the boundary of the tissue, $N_{2}$, and the number of all other spring-type vertices, $N_{3}$. These are illustrated in Fig. \ref{fig_models_coordination_number_penetration_depth} (a). Note that $N_1 = N_2$ ($N_1$ is not equal to the number of total vertices on the tissue boundary).

The number of total springs in the spring network, $M$, is given by
\begin{align}
    M &= N_{2} + \frac{1}{2} (2 N_{2}) + \frac{1}{2} (3 N_{3})
    \label{eq_Nsprings_rigid_boundary_01_line_1}
    \\
      &= 2 N_{2} + \frac{3}{2} N_{3}
    \label{eq_Nsprings_rigid_boundary_01_line_2}
\end{align}
I.e., each of the $N_{2}$ vertices has three springs connected to it-- one which is connected to the tissue boundary and only gets counted once (first term in Eq. \ref{eq_Nsprings_rigid_boundary_01_line_1}) and two which are connected to other spring-type vertices and will get counted twice (second term in Eq. \ref{eq_Nsprings_rigid_boundary_01_line_1}) when we add the three springs associated with all $N_{3}$ vertices (third term). Hence the factors of $1/2$ in the second and third terms.

If we assume that the tissue is highly rigid, the tissue-ECM interface acts as a fixed boundary for the spring network. This means that although there are springs connected to the boundary, the $N_{1}$ vertices on the boundary do not contribute to the network degrees of freedom. The total number of vertices we consider as part of the network in that case is
\begin{equation}
    N_{total, rigid} = N_{2} + N_{3}
 \label{eq_Nvertices_rigid_boundary_01}
\end{equation}
Substituting our expressions for $M$ and $N_{total, rigid}$ for $N_{edges}$ and $N_{vertices}$ in Eq. \ref{eq_avg_coord_num_01}, respectively, we find
\begin{equation}
    \langle z \rangle = \frac{4 N_{2} + 3 N_{3}}{N_{2} + N_{3}}.
    \label{eq_avg_coord_num_rigid_01}
\end{equation}
Noting that $N_{total, rigid} = N_{2} + N_{3}$ is also equal to the total number of spring-type vertices, we rewrite our expression for $\langle z \rangle$ in terms of the total number of spring-type vertices, $N_{v., sp., all}$, and ultimately in terms of the fraction of spring-type vertices on the boundary to total spring-type vertices, $f_{boundSp/allSp}$:
\begin{equation}
    \langle z \rangle = 3 + f_{boundSp/allSp}.
    \label{eq_avg_coord_num_rigid_final}
\end{equation}
Since $f_{boundSp/allSp} \geq 0$, in the case of a rigid boundary, $\langle z \rangle \geq 3$.



Based on the reasoning used here, we predict that embedding a fluid-like tissue, as opposed to a rigid one, will decrease $l^*_{0}$. This is because adding a fluid-like tissue will create a floppy boundary, increasing the number of spring network degrees of freedom without changing the number of springs, and effectively lowering the coordination number.

\begin{figure}
    \centering
    \includegraphics[width=0.32\textwidth]{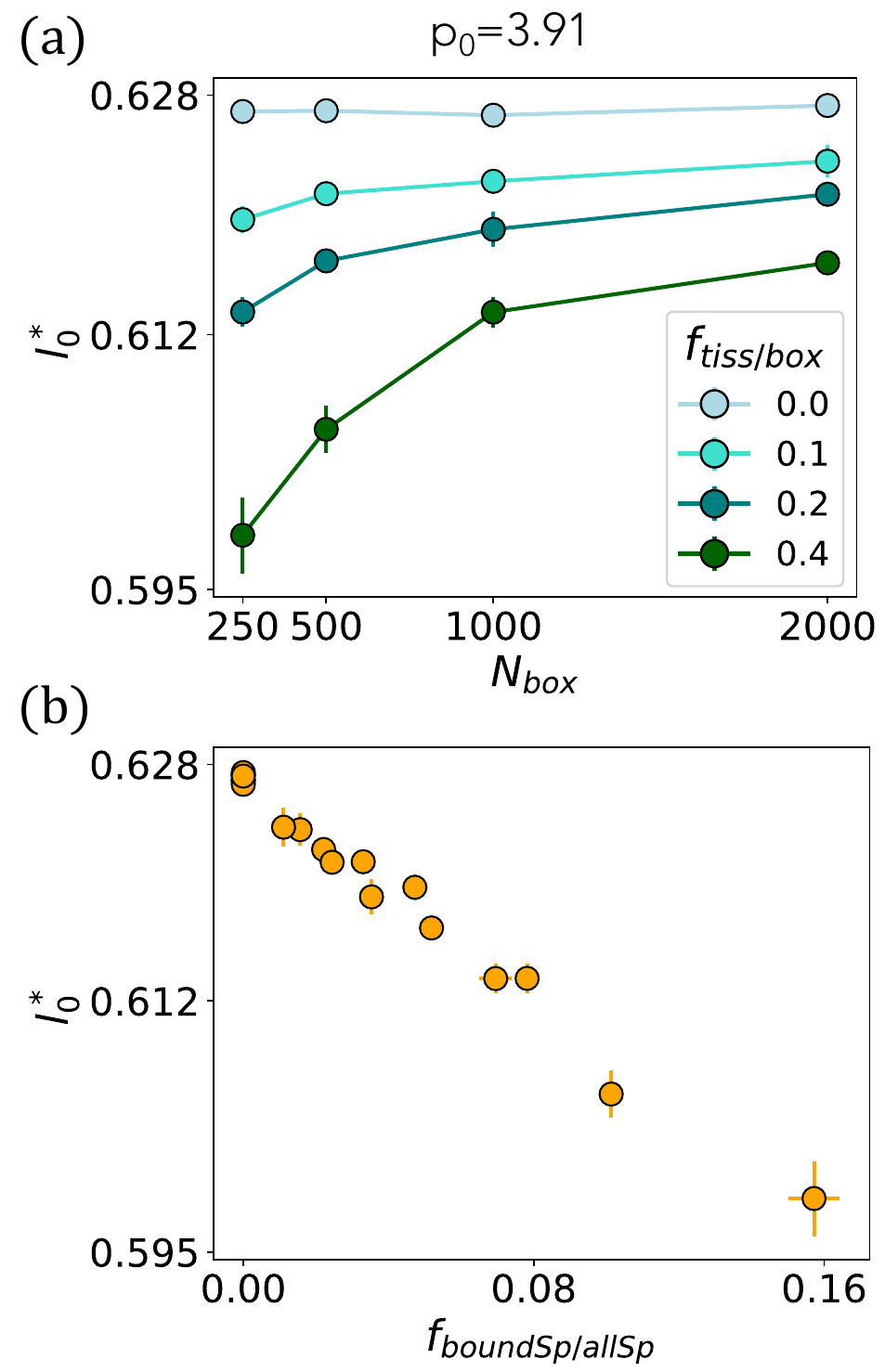}
    \caption{\textit{The dependence of $l^*_{0}$ on system size and tissue fraction, for a tissue with $p_0=3.91$.} (a) $l^*_{0}$ is plotted against the total system size for four values of the percentage of the box taken up by tissue. (b) For each data point in (a), we calculate the number of spring degrees of freedom connected to the boundary and plot this value (normalized by the total number of spring degrees of freedom) vs. $l^*_{0}$.}
    \label{fig_p0_3p91_l0star_boxsize_fraction_DOF}
\end{figure}

To test this, we repeat our analysis of $l^*_{0}$ as a function of system size and tissue fraction, this time for an embedded tissue with $p_{0}=3.91$. The results are shown in Fig. \ref{fig_p0_3p91_l0star_boxsize_fraction_DOF}, where we see that for a network with fluid-like embedded tissue, where $f_{boundSp/allSp} > 0$, $l^*_{0}$ is always less than that of a spring network alone, as we predict.

{}

We also repeat our constraint-counting analysis, now including the vertices on the tissue-ECM boundary as contributing to the spring network's degrees of freedom. In this case, we have the same number of total springs,
\begin{equation}
    M = 2 N_{2} + \frac{3}{2} N_{3}, 
 \label{eq_Nsprings_floppy_boundary_01}
\end{equation}
but the number of vertices that contribute to the degrees of freedom is now
\begin{equation}
    N_{total, floppy} = N_{1} + N_{2} + N_{3}.
 \label{eq_Nvertices_floppy_boundary_01}
\end{equation}
Because $N_{1}=N_{2}$, this can be written as
\begin{equation}
    N_{total, floppy} = 2 N_{2} + N_{3}.
 \label{eq_Nvertices_floppy_boundary_02}
\end{equation}
Putting Eq. \ref{eq_Nsprings_floppy_boundary_01} and Eq. \ref{eq_Nvertices_floppy_boundary_01} into Eq. \ref{eq_avg_coord_num_01} results in
\begin{equation}
    \langle z \rangle = \frac{4 N_{2} + 3 N_{3}}{2 N_{2} + N_{3}}.
 \label{eq_avg_coord_num_floppy_01}
\end{equation}
Again, we can replace $N_{2} + N_{3}$ with $N_{v., sp., all}$, and in doing so, we find
\begin{equation}
    \langle z \rangle = \frac{3 + f_{boundSp/allSp}}{1 + f_{boundSp/allSp}}.
 \label{eq_avg_coord_num_floppy_final}
\end{equation}
For $f_{boundSp/allSp} \geq 0$, which is always the case, $\langle z \rangle \leq 3$. 

\begin{figure}
    \centering
    \includegraphics[width=0.35\textwidth]{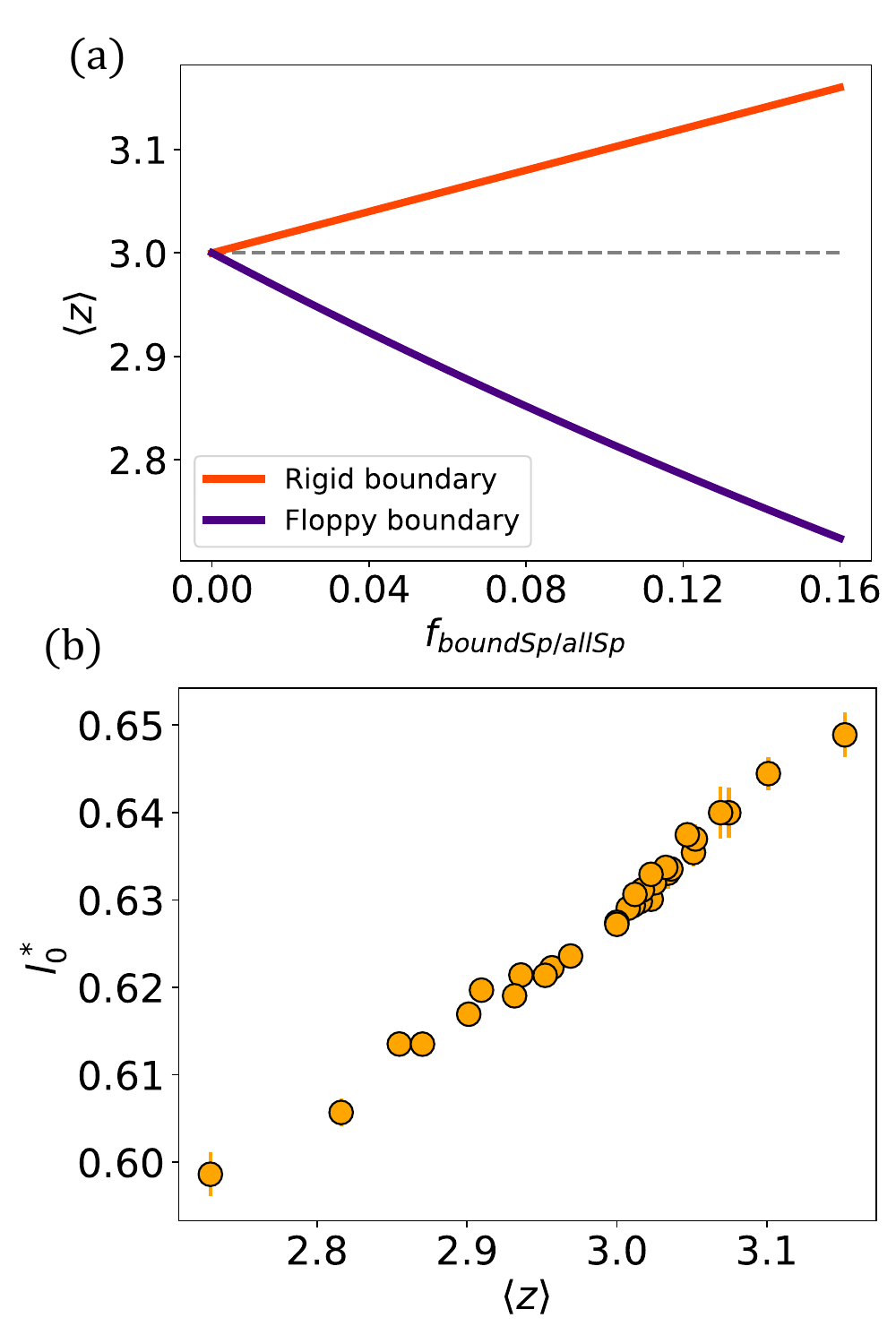}
    \caption{\textit{Effective coordination number.} (a) The theoretical relationships between the fraction of boundary-connected, spring network vertices to total spring network vertices and effective coordination number, for a rigid boundary (Eq. \ref{eq_avg_coord_num_rigid_final}) and floppy boundary (Eq. \ref{eq_avg_coord_num_floppy_final}), are plotted. The value $\langle z \rangle = 3$ is plotted as a dotted line for reference. (b) Using the relationships in (a), the data in Fig. \ref{fig_p0_3p71_l0star_boxsize_fraction_DOF} (b) and Fig. \ref{fig_p0_3p91_l0star_boxsize_fraction_DOF} (b) are plotted together against $\langle z \rangle$.}
    \label{fig_avgZ_vs_boundFrac_l0_theory_data}
\end{figure}

We can now directly compare the effective coordination number for a spring network with embedded floppy tissue, Eq. \ref{eq_avg_coord_num_floppy_final}, to that of a network with rigid tissue, Eq. \ref{eq_avg_coord_num_rigid_final}. This comparison is shown in Fig. \ref{fig_avgZ_vs_boundFrac_l0_theory_data} (a). In addition, using these relationships between $f_{boundSp/allSp}$ and $\langle z \rangle$, we now plot $l^*_{0}$ against $\langle z \rangle$. In Fig. \ref{fig_avgZ_vs_boundFrac_l0_theory_data} (b), we replot the data in Figs. \ref{fig_p0_3p71_l0star_boxsize_fraction_DOF} and \ref{fig_p0_3p91_l0star_boxsize_fraction_DOF} this way. 

Despite the simple, mean-field arguments used to extract an effective coordination number for each configuration, and the fact that we do not expect constraint-counting to capture all aspects of rigidity in these systems, our data collapses reasonably well across the system sizes, tissue fractions and tissue rigidities tested.

{}

\subsection{Transition behavior and mechanism}  \label{sec_ecm_diagram_sharpness}

We have shown that the shift in the spring network's transition point depends on the relative size and fluidity of the embedded tissue. A fluid-like tissue, for example, with a relatively large boundary, lowers the effective average coordination number of the surrounding spring network, requiring a larger strain, and therefore a smaller $l_0^*$, to rigidify it. 

For a pure spring network, it has been shown that at the point of rigidification, the springs can no longer all achieve their preferred lengths, due to a geometric incompatibility. Due to the network connectivity and geometry, there are suddenly no accessible states in which all of the springs have zero energy \cite{Merkel_2019, Arzash_2020}. Is this also what drives the onset of spring network rigidity in the presence of an embedded tissue? If so, how does the interaction with the tissue help or hinder the network in finding zero-energy states when under strain?

For example, based on our results, we hypothesize that a fluid-like tissue allows the network to remain ``compatible'' with the imposed strain at lower $l_0$ values than if there were no embedded tissue-- hence the smaller $l_0^*$. But what happens, then, at the point that the network does become rigid?

It has previously been observed, in both two-dimensional spring network models and vertex models, that at the moment the system becomes rigid, it sustains a single, one-dimensional state of self-stress that spans the network \cite{Merkel_2019}. In our composite system, we might expect this to manifest as a chain of rigid cells through the tissue bulk. Instead, at the moment of network rigidification, we see a layer of cells on the tissue boundary become rigid, forming a rigid loop around a still-floppy tissue bulk.

\begin{figure}
    \centering
    \includegraphics[width=0.49\textwidth]{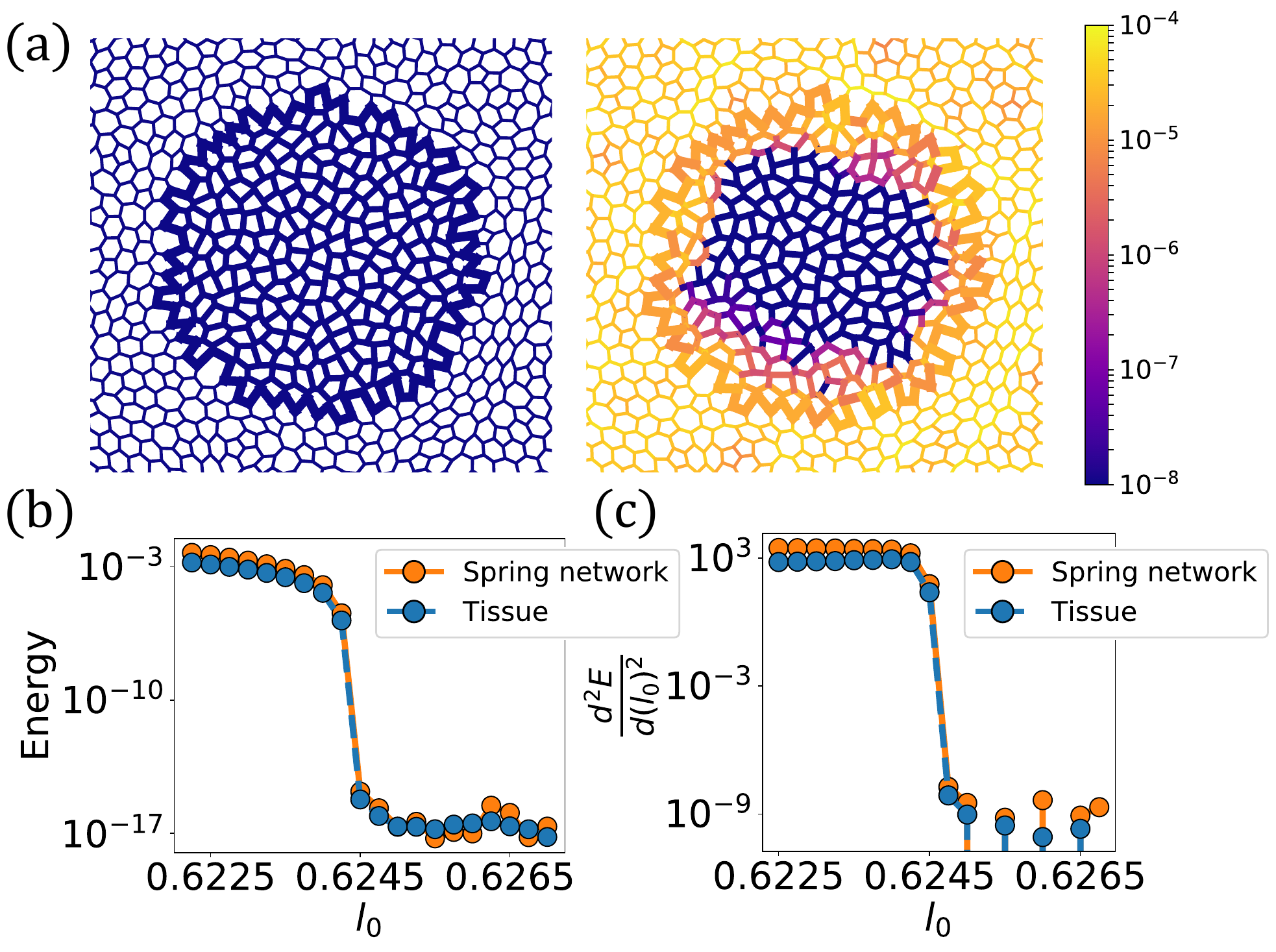}
    \caption{\textit{Onset of network rigidity: embedded tissue with $p_0 = 3.91$.} (a) Snapshots of a system configuration before and after the onset of spring network rigidification (we have cropped out much of the spring network to better observe the boundary). Springs and cell edges are colored by their tensions (see colorbar). Cell edges on the tissue boundary are shown in the thickest lines. While the network is floppy, both the tissue and ECM edges have effectively zero tension. At the moment the network becomes rigid, the springs' tensions and a subset of cell edge tensions along the tissue boundary increase substantially. (b) The total energies of the spring network and tissue are plotted against $l_0$. (c) The second derivative of the energies of the spring network and tissue are plotted against $l_0$.}
    \label{fig_rigidity_onset_floppy_tissue}
\end{figure}{}

This phenomenon is illustrated in Fig. \ref{fig_rigidity_onset_floppy_tissue} (a), where we have colored all edges by their tensions, including tissue-type edges using the definition given earlier. In this example, $p_0=3.91$, and on the left, the entire system-- tissue and network-- is floppy and all tensions are effectively zero. However, as we cross the transition point, the network and tissue exterior become rigid, while the tissue bulk remains floppy (right). We see this rigidification quantitatively by plotting the energies of the spring network and tissue as a function of $l_0$ (Fig. \ref{fig_rigidity_onset_floppy_tissue} (b)) and the second derivative of the energies with respect to $l_0$ (Fig. \ref{fig_rigidity_onset_floppy_tissue} (c)).

\begin{figure}
    \centering
    \includegraphics[width=0.42\textwidth]{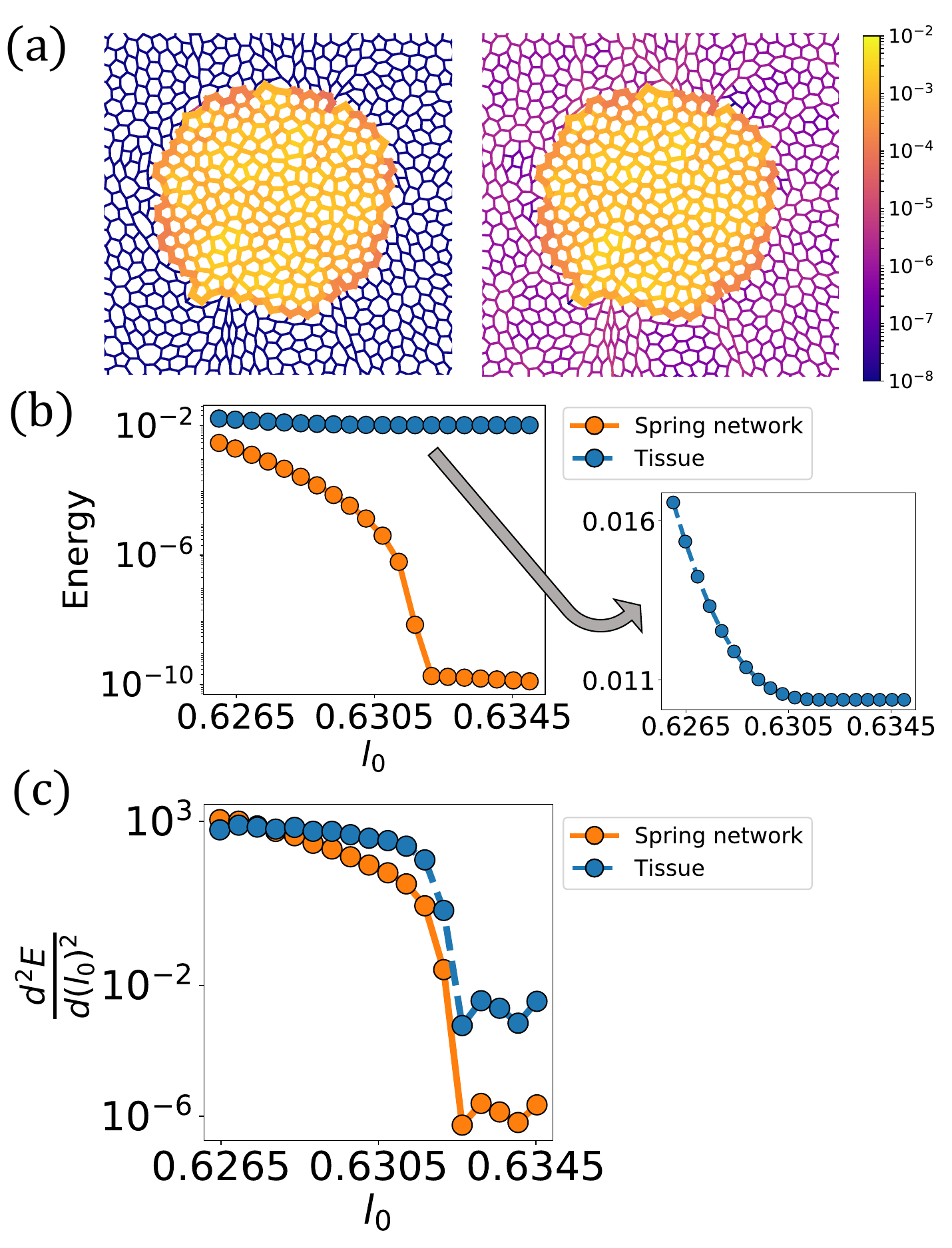}
    \caption{\textit{Onset of network rigidity: embedded tissue with $p_0 = 3.71$.} (a) Snapshots of a system configuration before and after the onset of spring network rigidification (we have cropped out much of the spring network to better observe the boundary). Springs and cell edges are colored by their tensions (see colorbar). Cell edges on the tissue boundary are shown in the thickest lines. While the network is floppy, the tissue cell edges have finite tension, since the tissue is rigid-like and therefore geometrically frustrated. (b) The total energies of the spring network and tissue are plotted against $l_0$. The tissue energy does also increase at the onset of network rigidity (detail). (c) The second derivative of the energies of the spring network and tissue are plotted against $l_0$.}
    \label{fig_rigidity_onset_rigid_tissue}
\end{figure}{}

To gain physical insight into why this ring-like configuration might be energetically-favorable, we model the tissue as a symmetric, regular lattice of seven hexagons, as shown in Fig. \ref{fig_models_coordination_number_penetration_depth} (b). This system represents the tissue, with each hexagon representing a cell. We imagine that each of the protruding vertices on the outer perimeter (blue vertices) are coupled to springs, which also lie on a honeycomb lattice. (The black vertices on the perimeter are already three-fold coordinated and do not couple to springs). The inner hexagon (darker blue) is the only hexagon not coupled to springs and represents the tissue bulk.

We use the first sum in Eq. \ref{eq_nondim_energy_total} as our Hamiltonian for the tissue, and we assume that the preferred area and perimeter for each cell are those of a regular hexagon, $a_{0}=(3\sqrt{3}/2) l^2$ and $p_{0}=6l$, and let $l=1$. If the protruding outer vertices are all displaced by an amount, $d_{0}$, along the springs to which they are coupled, what is the energetically-favorable amount, $d_{1}$, for the inner vertices to be displaced? In other words, if there is tension in the spring network, and the outer vertices are therefore being pulled, is it helpful, energetically, for vertices in the bulk to also move? 

We find that for $d_{0} > 0$, $0 < d_{1} < d_{0}$, meaning that it is favorable for the inner vertices to move, but not as much as the outer vertices. Specifically, the relationship between $d_1$ and $d_0$ is well-fit by $d_1 = A (d_{0})^B$, where $A=0.328$ and $B=1.062$. Assuming subsequent layers follow this same relationship (i.e. $d_N = A (d_{N-1})^B$), we find that the vertex displacement falls off quickly as $d_N = (A^{B^{N-1} + B^{N-2} + \ldots + B^0}) ((d_{0})^{B^N})$. Although we do not expect quantitative agreement between our simplified model and numerical observations, this model suggests a physical reason for the observed ``rigid ring'' seen in our simulations: it is energetically-favorable to have a sharp fall-off in rigidity towards the center of the tissue spheroid when pulling radially on a roughly circular, three-fold coordinated vertex model region. Interestingly, our findings align with more recent three-dimensional vertex modeling results reporting a boundary-bulk effect in which the boundary cells of a tissue spheroid are organized differently than the remaining bulk cells~\cite{Zhang2023}. 

In this toy model, we assume the tissue starts in a zero-energy state before the outer vertices are displaced. However, for a tissue that is already frustrated, we still expect an increase in frustration on the boundary when pulled on by springs, and therefore a fall-off of this additional frustration as we approach the tissue bulk. This is because frustrated tissue is solid-like, with an average cell shape index above the preferred value, meaning that the cells are less circular than they would like to be, energetically speaking. Pulling radially on the tissue only further elongates the cells on the boundary, driving them even further from their preferred shapes and increasing their frustration. This is illustrated in Fig. \ref{fig_rigidity_onset_rigid_tissue}, where in (a) we have again plotted configuration snapshots above and below the spring network's rigidity transition, this time with embedded tissue with $p_0  = 3.71$. In (b) and (c), both the tissue and spring network's total energies and second derivatives of energy are plotted against $l_0$. Given the relative scales of network and tissue tensions in this case, it is difficult to see, by eye, the effect on the tissue at the onset of network rigidity. However, looking at the tissue's energy alone (detail, Fig. \ref{fig_rigidity_onset_rigid_tissue} (b)), we see that the tissue does respond to the increase in network tension. This response is highlighted in (c). The tissue therefore undergoes a cross-over from a less rigid to more rigid state at the same time that the network goes from floppy to rigid.

However, despite the similarities in the physical response of solid and fluid tissues to rigidifying networks, Figs. \ref{fig_rigidity_onset_floppy_tissue} (b) and \ref{fig_rigidity_onset_rigid_tissue} (b) indicate that the types of transitions may differ. For fluid-like embedded tissue, the network's energy seems to jump from zero to non-zero, while the energy of a network with solid-like embedded tissue changes more gradually. To explore this, we compare the average network tension and participation ratio, $\psi$, as a function of $l_0$, between networks with embedded tissue of $p_0=3.71$ and $p_0=3.91$. In each case, we study three system sizes and average over 50 simulations with different initial configurations. All simulations have $f_{tiss/box} = 0.1$.

\begin{figure}
    \centering
    \includegraphics[width=0.49\textwidth]{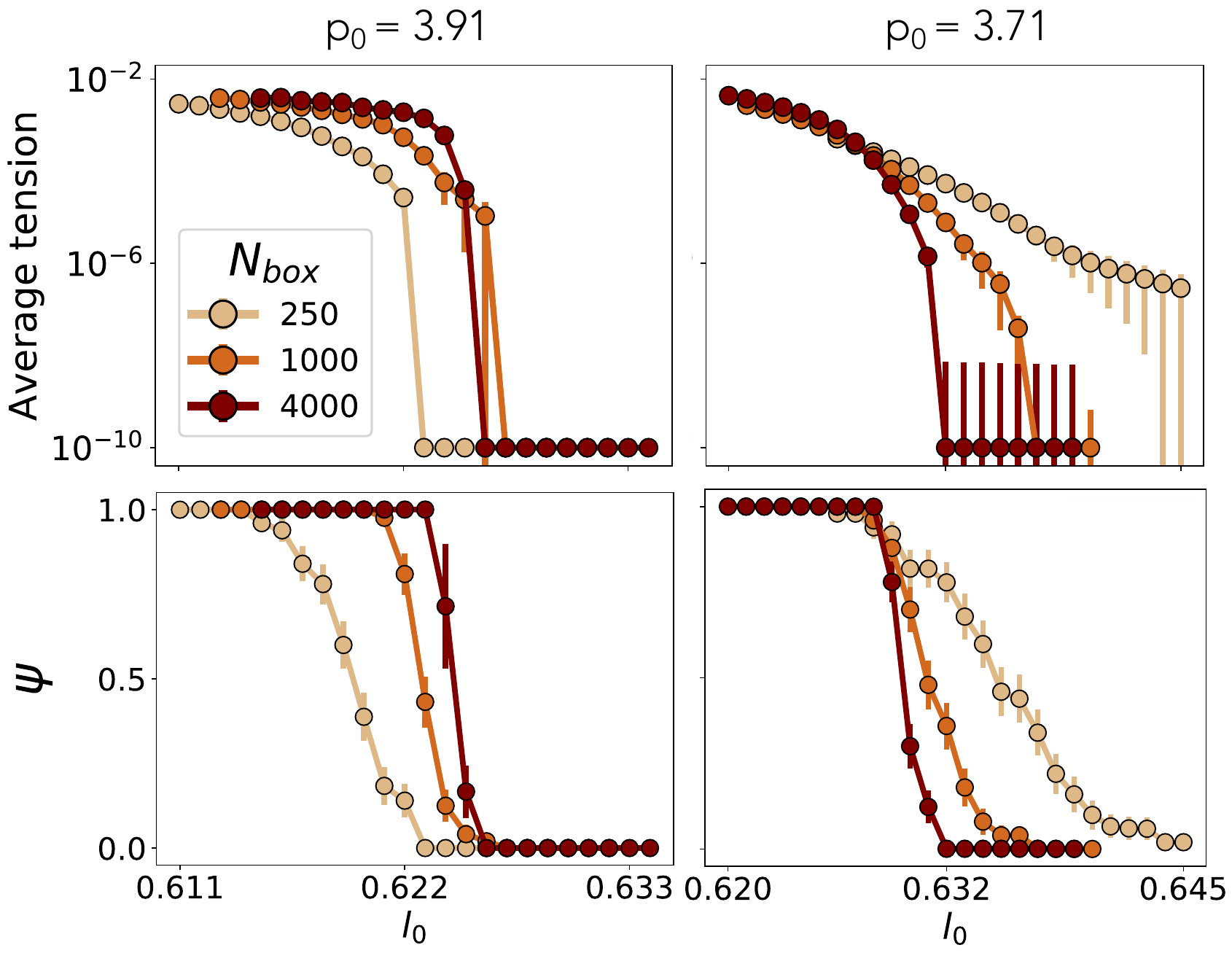}
    \caption{\textit{Network transition behavior.} The average network tension (top row) and participation ratio, $\psi$, (bottom row) are shown across three system sizes for fluid-like (left column) and solid-like (right column) embedded tissue. In all cases, $f_{tiss/box}=0.1$.}
    \label{fig_transition_sharpness_analysis}
\end{figure}{}

The results are shown in Fig. \ref{fig_transition_sharpness_analysis}. Looking first at the bottom row, we see that $\psi$ is affected similarly across $p_0$ values as a function of system size. For a fluid-like tissue ($p_0=3.91$), the inflection point, which we use to define $l_0^*$, increases with increasing system size, as observed in Fig. \ref{fig_p0_3p91_l0star_boxsize_fraction_DOF}. What we could not see in that figure was the simultaneous increasing sharpness of the curve. For a solid-like embedded tissue ($p_0=3.71$), the trend is similar, except for the decreasing position of the inflection point. 

The participation ratio is simply a measure of the number of springs above a certain tension threshold, as defined earlier. Other than being above the threshold, $\psi$ does not hold any further information about the magnitudes of the spring tensions. Therefore we also studied the average spring network tension (Fig. \ref{fig_transition_sharpness_analysis}, top row). For $p_0=3.71$, there is a clear trend with system size, from a very smooth and gradual increase in tension with decreasing $l_0$ at $N_{box} = 250$, to a well-defined transition at $N_{box} = 4000$. On the other hand, for $p_0=3.91$, there is no such obvious trend. The behavior across system sizes remains sharp and well-defined, and it quickly-converges (the curves for $N_{box} = 1000$ and $N_{box} = 4000$ are similar). However, from the trends in both participation ratio and tension, across $p_0$ values, the gap at the transition point seems to be growing with increasing system size, suggesting a discontinuous transition, as observed previously for bulk spring networks \cite{Vermeulen2017}.

\subsection{Effects of interfacial tension} \label{sec_interfacial_tension}

Until now, we have ignored any effect of interfacial tension on the surface of the tissue spheroid. However, interfacial tension is an important characteristic of cell aggregates that may play a role in whether tissue cells migrate away from the tissue bulk \cite{Yousafzai_2020, Beaune_2018, Hall_2016,Zhang2024}. We therefore now add a term to our Hamiltonian (Eq. \ref{eq_nondim_energy_total}) which acts to modulate the tension the tissue boundary edges:

\begin{multline}
    e_{total} = \sum_{cells, \alpha} \Big[(a_{\alpha} - 1)^{2} + k_{p} (p_{\alpha} - p_{0})^{2} \Big]\\ 
     + \frac{1}{2} k_{sp} \sum_{sp,ij}  (l_{ij } - l_{0})^{2}
     + \gamma \sum_{int,ij } l_{ij}
 \label{eq_nondim_energy_total_tension}
 \end{multline}
In the last term, $\gamma$ is the line tension on each interface edge and the sum is over all interface edges.
 
 \begin{figure}
    \centering
    \includegraphics[width=0.49\textwidth]{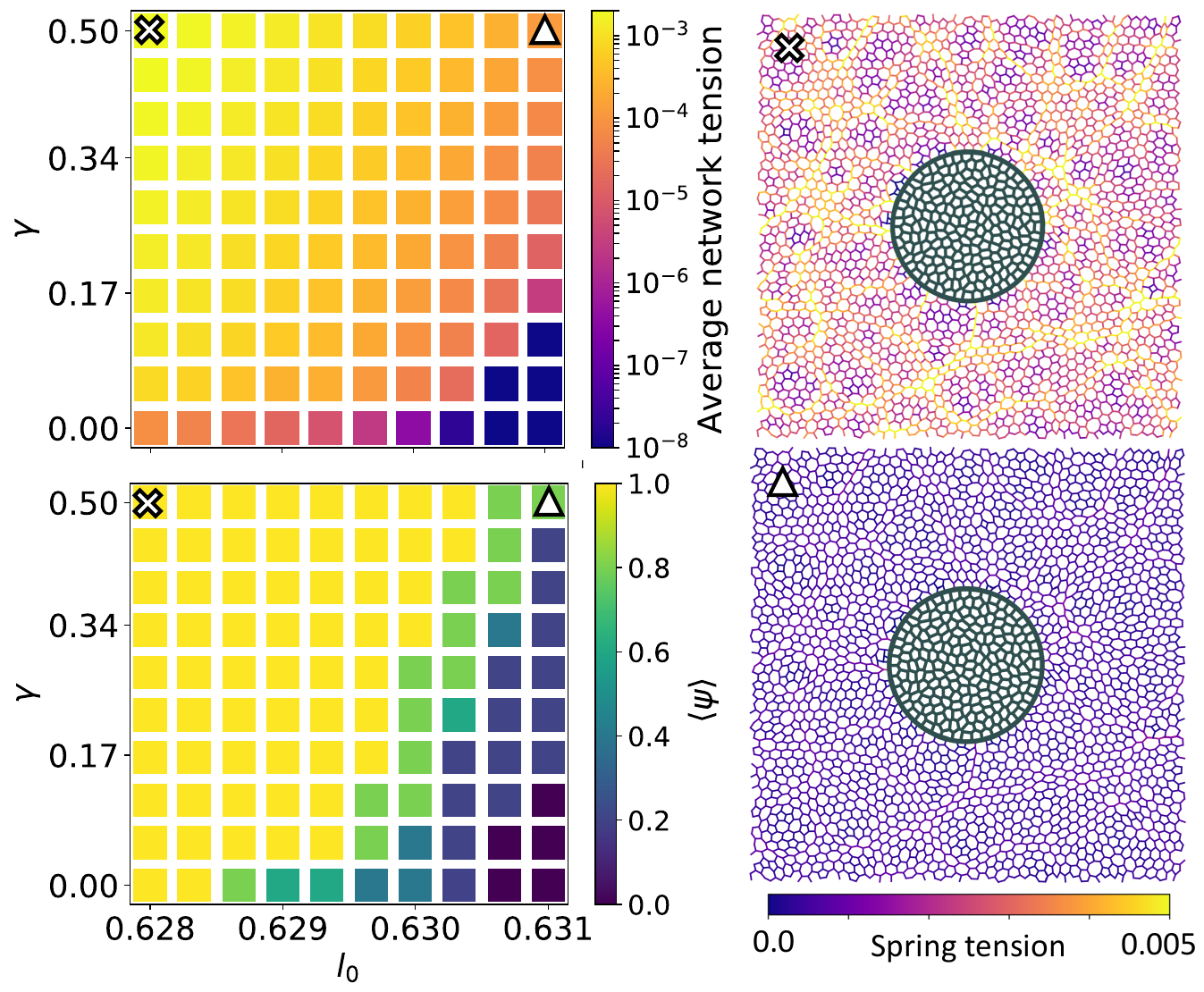}
    \caption{\textit{The effect of interfacial tension.} A phase diagram for the spring network, quantified by the participation ratio of tense springs, $\psi$ (averaged over configurations), is shown as a function of the preferred spring length, $l_0$, and tissue interfacial tension, $\gamma$. In this example, $N_{box} = 2000$, $f_{tiss/box} = 0.1$, and $p_0 = 3.71$.}
    \label{fig_l0_vs_gamma_p0_3p7100}
\end{figure}{}
 
 Increasing $\gamma$ results in the tissue becoming highly circular with decreasing radius. This means that the tissue can act as a contractile inclusion, deforming the network to which it is attached. In this way, floppy networks can be stiffened via an increase in tissue interfacial tension, as shown in Fig. \ref{fig_l0_vs_gamma_p0_3p7100} with $p_0=3.71$, $N_{box}=2000$, and $f_{tiss/box}=0.1$. In the left column, the average network tension and average participation ratio, $\psi$, are shown as a function of the tissue interfacial tension, $\gamma$, and the spring network's preferred edge length, $l_0$. At $\gamma=0$, we see a somewhat smooth transition from floppy to rigid at $l_0 \approx 0.629$. As $\gamma$ increases, the transition point shifts to the right, meaning that the spring network is rigid at higher values of $l_0$. For values of $l_0$ at which the network would ``normally'' be floppy, with sufficient tissue interfacial tension, the network is rigid. In the right column, we show snapshots of two minimum-energy configurations corresponding to the parameter values indicated by the Xs and triangles in the phase diagrams. Here we are only coloring the springs by their tension and leaving the tissue cell edges uncolored.
 


\section{Discussion} \label{sec_discussion}

Using a vertex model to represent a tissue spheroid and a spring network to represent the surrounding extracellular matrix, we have studied the effects of a tissue spheroid on the rheology of its spring network environment. These effects are a function of the relative size of the tissue spheriod and spring network, as well as the tissue spheroid's own emergent rheology. The latter point emphasizes the importance of utilizing a cellular-based model capturing a rigidity transition that is observed in tissues~\cite{Park_2015,Mongera_2018}. Specifically, we find that the size and fluidity of the tissue spheroid can be related to a new, effective coordination number for the network, which can be related to the network's phase diagram. Higher effective coordination numbers result in stiffer networks for the tuning parameter, $l_0$, while lower numbers result in floppier networks. This suggests that for \textit{in vitro} experiments, where tissue spheroids are embedded in collagen matrices, the relative size of spheroids may play a role in determining the matrix rigidity and subsequently spheroid behavior, such as migration propensity and morphology, though including three-body interactions into the spring network will presumably reduce this role with more mechanical constraints in the network itself~\cite{Quint2011,Broedersz_2014}.

We also find that the onset of rigidity in the spring network is accompanied by a ring of rigidity along the tissue boundary. Because our tissue is represented by a vertex model and, therefore, itself a composite of individual cells, we observe that the cells become less solid-like cells as a function of distance from the boundary, i.e., the cells near the core of the spheroid are less solid-like than the boundary. This may help to explain one of our previous results of this model that cells being pulled by the surrounding network, though elongated and ostensibly primed for migration, are solid-like, due to induced cell-cell alignment \cite{Parker_2020}. It is now clear that for the network to be rigid, a portion of the tissue must also be rigid, and that it is energetically-favorable for this portion to be a ring on the exterior of the tissue. Finally, adding interfacial tension to the tissue boundary enables the tissue to act as a contractile body, applying stress to an otherwise-floppy network.

Altogether, our results suggest that changes to the network phase diagram due to an embedded tissue spheroid are ultimately dominated by the deformability and relative length of the tissue boundary. As the computational modeling of embedded spheroids becomes even more complex, including an aggregate of cells that actively pulls on its surroundings~\cite{Zhang2024}, one can imagine more explicit biomechanical feedback between cells and ECM to ultimately determine how cells break away from tissue spheroids. 

The authors acknowledge Cristina Marchetti and Lisa Manning for discussion during early stages of this project. JMS acknowledges financial support from the National Science Foundation via grants PoLS-2014192 and PolS-2412961.


\bibliographystyle{unsrt}
\bibliography{ref}

\end{document}